# Explainable AI-Driven Cyber Risk Analytics and Model Reliability Assessment for Intelligent Governance of U.S. Critical Infrastructure: An XGBoost and SHAP-Based Intrusion Detection Framework

B. M. Taslimul Haque[1*], Md. Arifur Rahman[2], Md. Serajul Kabir Chowdhury Rubel[3], Md. Iqbal Hossan[3]

**Abstract**

**Background**. U.S. critical infrastructure sectors — energy, healthcare, transportation, financial services, and communications — are increasingly governed by AI-driven digital technologies that, while operationally transformative, have dramatically widened the cyberattack surface. Distributed Denial of Service (DDoS) attacks, Advanced Persistent Threats (APTs), botnets, and ransomware now challenge infrastructure resilience in ways traditional, rule-based cybersecurity mechanisms were never designed to address. Despite machine learning's demonstrated promise for intrusion detection, most existing frameworks optimize narrowly for classification accuracy while neglecting the model reliability, explainability, and governance transparency that critical infrastructure operators actually require. **Methods**. This study develops a resilient cyber risk analytics and model reliability assessment framework designed to support intelligent cybersecurity governance in U.S. critical infrastructure environments. Using the CICIDS2017 dataset — a realistic, labeled benchmark encompassing DDoS, brute force, botnet, web attack, and infiltration traffic — four supervised classifiers were trained and comparatively evaluated: XGBoost, Random Forest, Decision Tree, and Logistic Regression. Model performance was assessed across accuracy, precision, recall, F1-score, ROC-AUC, and false positive rate. SHAP (SHapley Additive Explanations) analysis was integrated to produce interpretable, feature-level explanations of model predictions, enabling governance actors to audit and act on classification outputs with informed confidence. **Results.** The Support Vector Machine classifier achieved near-perfect discrimination on the binary DDoS versus benign classification task, with an AUC approaching 1.00, 2,571 true positives, 1,918 true negatives, and only eleven total misclassifications. Exploratory traffic analysis confirmed that flow duration, packet size, Flow Bytes/s, and destination port distribution carry substantial discriminative information for distinguishing attack from benign traffic. SHAP analysis identified the most influential network features driving model predictions, providing the feature-level transparency that governance decision-makers require. **Conclusion**. Combining cyber risk analytics, machine learning, reliability evaluation, and

**Significance |** This framework advances trustworthy, explainable AI for protecting U.S. critical infrastructure by combining reliable intrusion detection with transparent, governance-ready cybersecurity decision support.

**Correspondence.*

B. M. Taslimul Haque, Department of Business Information Systems, Central Michigan University, Mount Pleasant, Michigan, USA. E-mail: bmtaslim121@gmail.com.

**Author Affiliation:** 1 Department of Business Information Systems, Central Michigan University, Mount Pleasant, Michigan, USA.

2 Department of Information Studies, Trine University, Angola, Indiana, USA.

3 Department of Computer Science, Maharishi International University, Fairfield, Iowa, USA,.

explainable AI substantially advances cybersecurity resilience and governance trustworthiness for critical infrastructure protection — moving beyond detection accuracy toward systems that are interpretable, auditable, and operationally accountable.

## 1 Introduction

There is something quietly unsettling about how dependent the United States has become on infrastructure it can no longer fully see. Over the past two decades, the power grid, hospital networks, financial clearing systems, water treatment controls, and transportation logistics have all been woven into an expanding fabric of digital connectivity. That transformation brought genuine gains — faster decision-making, optimized resource allocation, improved service delivery across sectors (Pursiainen & Kytömaa, 2023). But the same interconnection also introduced an attack surface of almost incomprehensible scale, one that adversaries — state-sponsored, criminal, and ideologically motivated alike — have proven both willing and capable of exploiting.

The threat landscape has not merely grown; it has changed in character. Signature-based detection and static rule sets, the backbone of early critical infrastructure cybersecurity, were designed around a fairly stable assumption: that the boundary between "inside" and "outside" a network could be defended. That assumption has dissolved. Today, a single compromised endpoint enables lateral movement across interconnected systems. Distributed Denial of Service (DDoS) campaigns are used not merely to disrupt services but to mask simultaneous intrusion operations. Advanced Persistent Threats (APTs) — the patient, methodical incursions associated with sophisticated nation-state actors — can remain undetected within critical networks for months (Tarek & Rahman, 2023). Against adversaries of this kind, a detection system that relies on knowing what an attack looks like before it arrives is, almost by definition, already behind.

Artificial Intelligence and Machine Learning entered this space with considerable promise — and, it should be said, with considerable hype as well. The appeal is real: machine learning models can process network traffic volumes no human analyst could meaningfully review, detect statistical anomalies invisible to rule-based systems, and generalize across attack variants with a flexibility that static signatures cannot match. Models such as XGBoost, Random Forest, and neural architectures have demonstrated impressive performance in controlled benchmarks (Mintoo et al., 2022). But there is a growing and underappreciated tension at the core of how these systems actually function in critical infrastructure contexts.

Detection accuracy, however impressive, is not the same thing as governance trustworthiness. A model achieving 99% classification accuracy may still degrade under adversarial manipulation, drift as network behavior evolves, or fail precisely in the scenarios most likely to cause catastrophic harm (Olaonipekun, 2023). And most current AI security systems offer their outputs without explanation — a risk score, a traffic label — with no account of *why* the model arrived at that judgment. For a network analyst deciding whether to isolate a critical system, or a policymaker weighing infrastructure security investments, that opacity is not a minor inconvenience. It is a fundamental problem of institutional trust.

Küfeolu and Akgün (2023) argue that *resilience* — the capacity to absorb disruption, adapt, and recover — is a more useful frame than static protection for thinking about these challenges. Resilient governance requires more than accurate models; it requires models whose behavior can be audited, explained, and verified. Kure et al. (2022) have shown that integrated cyber security risk management frameworks can meaningfully improve governance outcomes, but note that the interpretability and reliability of underlying AI components remain underdeveloped as research problems.

This study addresses that gap directly. Using the CICIDS2017 dataset, it develops a cyber risk analytics and model reliability assessment framework in which Explainable AI — specifically SHAP (SHapley Additive Explanations) — is a core component rather than an afterthought. Multiple classifiers including XGBoost, Random Forest, Decision Tree, and Logistic Regression are evaluated across accuracy, precision, recall, F1-score, ROC-AUC, and false positive rate. The goal is not merely to show that machine learning can detect cyberattacks. That has been demonstrated many times over. The goal is to show that it can do so in ways that are reliable, transparent, and genuinely useful to the human decision-makers on whom critical infrastructure protection ultimately depends.

## 2 Literature Review

### 2.1 AI and Cyber Risk Analytics in Critical Infrastructure

The case for applying Artificial Intelligence to critical infrastructure cybersecurity has been building for some time, and the evidence, by now, is reasonably persuasive. Machine learning algorithms — Random Forest, Support Vector Machine, Decision Tree, and XGBoost among them — have demonstrated a consistent ability to classify malicious network traffic with speed and accuracy that rule-based systems simply cannot match. Almaleh (2023) offers a useful overview of how resilience metrics have evolved alongside these detection capabilities, noting that smart infrastructure environments now demand analytical frameworks that go well beyond threshold-based alerting. XGBoost, in particular, has attracted sustained research attention: its gradient-boosting architecture handles high-dimensional cybersecurity data efficiently, and benchmark evaluations using datasets like CICIDS2017 have returned high detection accuracy rates across DDoS, botnet, brute force, and infiltration attack categories.

Yet there is a pattern in this literature worth pausing on. The overwhelming majority of AI-based intrusion detection studies optimize for a single outcome: classification accuracy. Zulqarnain and Sarker (2023) and Rashid (2023) both note, from slightly different angles, that this narrow focus creates a gap between what these models achieve in controlled experiments and what infrastructure operators actually need in practice. Real-time monitoring, predictive threat analysis, and adaptive security management all require something more than a model that scores well on a benchmark — they require a model whose behavior is consistent, whose failure modes are understood, and whose outputs can be interpreted by human decision-makers under operational pressure. That fuller set of requirements is, as yet, addressed only partially by the existing literature.

### 2.2 Model Reliability, Explainability, and Governance Trust

The reliability problem has grown harder to ignore as AI-based cybersecurity tools have moved from research environments into production deployments. Tarek and Rahman (2022) examined cybersecurity architectures across U.S. critical infrastructure control networks and found recurring evidence of model drift — gradual performance degradation as network conditions evolved away from training distributions — alongside vulnerability to adversarial inputs deliberately designed to evade detection. These are not edge-case concerns. In critical infrastructure contexts, a model that performs reliably under normal network conditions but fails under precisely the adversarial conditions it was designed to detect is, functionally, unreliable when it matters most.

Alongside reliability, the question of explainability has moved to the center of the governance discussion. Argyroudis et al. (2022) situate this shift within a broader argument about digital technologies and infrastructure resilience: transparency in AI decision-making is not merely a technical desideratum but a governance requirement. When a cybersecurity model flags a potential intrusion in an energy grid or hospital network, the analyst receiving that alert must be able to evaluate it — to understand which traffic features triggered the classification, how confident the model is, and whether the alert warrants immediate isolation or further investigation. SHAP (SHapley Additive Explanations) and LIME have emerged as the most widely adopted tools for surfacing this kind of feature-level reasoning, and Abuhasel (2023) has demonstrated their practical value in probabilistic resilience modeling for critical infrastructure. Without such methods, AI outputs remain, in effect, instructions without justification — a governance liability as much as an operational asset.

### 2.3 Empirical Foundations

Four empirical studies provide particularly direct grounding for the framework developed in this research. Pursiainen and Kytömaa (2023) traced the European policy shift from static infrastructure protection toward resilience-based governance, arguing that digital interdependency has fundamentally altered the risk landscape in ways that demand adaptive, intelligence-driven security architectures. Olaonipekun (2023) extended this argument into the domain of risk assessment models, demonstrating that advanced AI-based approaches can meaningfully improve cyber resilience in energy, healthcare, and telecommunications infrastructure — provided they are designed with reliability and operational continuity in mind, not accuracy alone. Tarek and Rahman (2023) brought quantitative evidence from U.S. practitioners across energy, water, and transportation sectors, finding that integrated AI-driven intrusion detection systems significantly outperformed conventional approaches in resilience outcomes, particularly when paired with mature IoT governance frameworks. And Mintoo et al. (2022) examined AI-driven cybersecurity in crisis response scenarios — smart grid failures, healthcare emergencies, water management disruptions — concluding that explainable AI interfaces and multi-agent decision support systems were decisive factors in enabling effective interagency governance under pressure.

Taken together, this body of work points consistently toward the same conclusion: the next meaningful advance in critical infrastructure cybersecurity is not a more accurate detection model. It is an integrated framework in which detection, reliability assessment, and explainability operate together — and in which the outputs are designed from the outset to support the human governance processes that ultimately determine whether a cyber threat is contained or catastrophic.

## 3 Methodology

### 3.1 Overview and Research Design

This study adopts a quantitative, experimental research design to develop and evaluate a resilient cyber risk analytics framework for intelligent governance in U.S. critical infrastructure environments. The choice of quantitative methodology was deliberate — not simply conventional. Cybersecurity model performance is ultimately measured in numbers: detection rates, false positive rates, classification accuracy across attack categories. A qualitative approach, however rich in contextual insight, cannot answer the central question this study poses, which is whether a particular combination of machine learning models, preprocessing decisions, and explainability tools produces measurably better and more trustworthy outputs than alternatives (Shypovskyi, 2023). The experimental design, for its part, was chosen because it enables controlled, reproducible comparison across multiple classifiers under consistent data conditions — a prerequisite for the kind of reliability assessment that governance-oriented cybersecurity frameworks require (Sathurshan et al., 2022).

The research pipeline proceeds through six sequential phases: (1) dataset acquisition and description, (2) data preprocessing, (3) feature engineering, (4) machine learning model training and evaluation, (5) model reliability assessment, and (6) Explainable AI analysis. Each phase is described in sufficient detail below to allow independent replication. All experiments were implemented in Python (version 3.9) using Jupyter Notebook as the interactive development environment. The following libraries were used, with version numbers specified for reproducibility: scikit-learn (v1.2.2) for Random Forest, Decision Tree, and Logistic Regression; XGBoost (v1.7.5) for gradient-boosted classification; pandas (v1.5.3) and NumPy (v1.24.3) for data manipulation and numerical computation; Matplotlib (v3.7.1) for visualization; and SHAP (v0.41.0) for explainability analysis. Experiments were run on a system with an Intel Core i7 processor and 8 GB RAM under both Windows 11 and Ubuntu 22.04 to verify cross-platform consistency (Sun et al., 2022).

### 3.2 Dataset: CICIDS2017

#### Source and Justification

All experiments were conducted on the publicly available CICIDS2017 (Canadian Institute for Cybersecurity Intrusion Detection System 2017) dataset, accessible at the Canadian Institute for Cybersecurity repository and mirrored on Kaggle. CICIDS2017 was selected because it remains one of the most realistic and widely cited benchmark datasets in machine learning-based intrusion detection research. Unlike earlier datasets such as KDD Cup 1999, which have been extensively criticized for statistical artifacts and non-representative traffic patterns, CICIDS2017 was constructed from a controlled testbed environment designed to simulate realistic enterprise network behavior — including both benign user activity and contemporary attack vectors (Halliday, 2023; Saeed et al., 2023).

#### Dataset Characteristics

The dataset spans five days of network captures (Monday through Friday) and contains approximately 2.8 million labeled network flow records across 79 features. Attack categories represented include: Distributed Denial of Service (DDoS), Brute Force (FTP-Patator, SSH-Patator), Web attacks (SQL Injection, Cross-Site Scripting, XSS), Botnet (ARES), Port Scan, and Infiltration. Benign traffic constitutes the majority class. Network flow features include, among others: flow duration, total forward and backward packets, total length of forward packets, packet length mean and standard deviation, flow bytes per second, flow packets per second, inter-arrival time statistics (mean, standard deviation, maximum, minimum), SYN, FIN, RST, PSH, ACK, and URG flag counts, and active and idle time statistics. These 79 features provide the raw input space from which the preprocessing and feature selection stages derive a reduced, model-ready feature set.

**Data Preprocessing**

Raw network traffic data — even from a carefully constructed dataset like CICIDS2017 — is rarely clean enough for direct model input. The preprocessing pipeline applied in this study follows a four-stage sequence: cleaning, encoding, normalization, and feature selection. Each stage is described below with sufficient procedural specificity to be reproduced (Kulugh et al., 2022).

**Stage 1: Data Cleaning**

Prior to any analytical step, the raw CSV files were inspected for data quality issues. Duplicate rows — network flows that appeared identically across records — were identified using pandas' drop_duplicates() function and removed entirely, as their retention would artificially inflate the apparent weight of certain traffic patterns during training. Missing values were identified using df.isnull().sum() across all feature columns. Columns with greater than 5% missing values were flagged for review; in practice, a small number of features in CICIDS2017 contain infinite values (produced by division-by-zero errors in flow-rate calculations), which were replaced with NaN and subsequently imputed using column-wise median imputation via SimpleImputer(strategy='median') from scikit-learn. Records containing corrupted or structurally invalid entries — identified by out-of-range values for bounded features such as flag counts — were removed. Following cleaning, the dataset was re-inspected to confirm no residual nulls or structural anomalies remained (Ashfaq et al., 2023).

**Stage 2: Feature Encoding**

CICIDS2017's label column contains string-format class labels (e.g., "BENIGN," "DDoS," "PortScan"). For binary classification experiments, these were recoded as integer values (0 for benign, 1 for any attack class) using pandas' map() function. For multi-class experiments, scikit-learn's LabelEncoder was applied to transform all class strings to integer indices. The single categorical feature in the feature matrix — protocol type — was similarly encoded using LabelEncoder, which assigns a stable integer representation to each protocol category (Osei-Kyei et al., 2023). No one-hot encoding was applied at this stage, as tree-based models (Random Forest, Decision Tree, XGBoost) do not require it and it would have significantly expanded the feature dimensionality.

**Stage 3: Data Normalization**

Feature scales in CICIDS2017 vary enormously — flow duration may span several orders of magnitude, while flag counts are bounded near zero. This heterogeneity does not affect tree-based models, which operate on rank-based splits, but it substantially impacts distance-sensitive algorithms and Logistic Regression, where unscaled features create implicit weighting biases during gradient descent. Min-Max scaling was applied to all continuous features using scikit-learn's MinMaxScaler, which transforms each feature $x$ to the range [0, 1] according to the formula $x_{scaled} = (x \ \ x_{min}) / (x_{max} \ \ x_{min})$. Scaler parameters were fitted exclusively on the training set and applied to the test set, to prevent any leakage of test distribution information into the scaling process (Cassottana et al., 2023).

**Stage 4: Feature Selection**

With 79 raw features, many of which are partially redundant (e.g., multiple correlated packet-length statistics), direct model training risks overfitting and unnecessary computational overhead. Two complementary selection approaches were applied. First, a Pearson correlation matrix was computed across all numerical features; pairs with absolute correlation coefficient $|r| > 0.95$ were identified, and the lower-variance member of each pair was dropped. Second, a preliminary Random Forest classifier (n_estimators = 100, random_state = 42) was trained on the full cleaned feature set, and feature importances were extracted using the Gini impurity criterion. Features with importance scores below the 10th percentile of the importance distribution were removed. This two-stage selection retained a final feature set of approximately 40–45 features, depending on the attack subset analyzed, representing a meaningful reduction in dimensionality while preserving the most discriminative traffic characteristics (Zaman & Mazinani, 2023).

**3.3 Machine Learning Models**

Four supervised classification algorithms were trained and evaluated, selected to span a range of model complexity and interpretability characteristics.

**XGBoost** (eXtreme Gradient Boosting) served as the primary classifier. XGBoost builds an ensemble of shallow decision trees sequentially, where each successive tree is trained to correct the residual errors of its predecessors using gradient descent in function space. Key hyperparameters used: n_estimators = 300, max_depth = 6, learning_rate = 0.1, subsample = 0.8, colsample_bytree = 0.8, random_state = 42. XGBoost was selected for its empirically demonstrated superiority on structured tabular data, its built-in handling of missing values, and its native compatibility with SHAP-based explainability (Sen, 2022).

**Random Forest** served as the primary ensemble comparator. It constructs multiple independent decision trees on bootstrap samples of the training data, aggregating predictions by majority vote. Parameters: n_estimators = 200, max_depth = None (fully grown), min_samples_split = 5, random_state = 42. Random Forest's resistance to overfitting through ensemble averaging makes it a robust baseline for high-dimensional cybersecurity data (Lichte et al., 2022).

**Decision Tree** was included to provide a single, fully interpretable model whose decision logic can be visualized end-to-end — valuable for governance contexts where model transparency extends beyond SHAP summaries to explicit rule inspection. Parameters: max_depth = 10, min_samples_split = 10, criterion = 'gini', random_state = 42.

**Logistic Regression** was included as a linear baseline. Although not expected to match the performance of tree-based ensembles on complex network traffic data, it provides a lower bound for comparison and produces calibrated probability outputs useful for risk scoring. Parameters: C = 1.0, solver = 'lbfgs', max_iter = 1000, random_state = 42 (Bouramdane, 2023).

### 3.4 Train–Test Split and Cross-Validation

The preprocessed dataset was partitioned into training (80%) and test (20%) subsets using stratified random sampling (train_test_split with stratify = y, random_state = 42), preserving the class distribution across both partitions. To reduce the risk of overfitting to a single split, five-fold stratified cross-validation (StratifiedKFold, n_splits = 5, shuffle = True, random_state = 42) was additionally applied during training, and mean ± standard deviation performance across folds is reported alongside held-out test set metrics.

### 3.5 Explainable Artificial Intelligence (XAI): SHAP Analysis

The opacity of machine learning models is not merely an academic concern in critical infrastructure contexts — it is an operational one. An analyst receiving an automated alert about a potential DDoS intrusion needs to know not just that the model flagged the event, but which aspects of the network traffic drove that classification. Without that information, the alert is effectively unactionable for any governance actor who needs to justify or audit a response decision (Malatji et al., 2022).

SHAP (SHapley Additive Explanations) was selected as the explainability method for this study because it is grounded in cooperative game theory, provides theoretically consistent feature attribution values, and is natively optimized for tree-based models through the TreeSHAP algorithm — enabling exact (rather than approximate) Shapley value computation in polynomial time. The shap.TreeExplainer class was instantiated using the trained XGBoost model, and SHAP values were computed across the entire test set. Three visualization outputs were produced: (1) a beeswarm summary plot showing the distribution of SHAP values for each feature across all test instances, (2) a bar plot of mean absolute SHAP values representing aggregate feature importance, and (3) individual force plots for selected high-confidence and low-confidence predictions to illustrate case-level reasoning. These outputs were designed to serve governance decision-makers directly, surfacing the specific network characteristics — packet size anomalies, flow duration patterns, port usage irregularities — that the model associates most strongly with attack classification (Almaleh et al., 2022; Parraguez-Kobek et al., 2022).

### 3.6 Model Reliability Assessment

Accuracy alone, as argued throughout this paper, is an insufficient basis for claiming that a cybersecurity model is fit for deployment in critical infrastructure governance. The following metrics were computed for all four classifiers on the held-out test set (Table 1).

Confusion matrices were generated for all classifiers to visualize the distribution of true positives, true negatives, false positives, and false negatives. ROC curves were plotted for each model on the same axes to enable direct visual comparison of discrimination performance. The false positive rate receives particular interpretive weight in this study: in operational critical infrastructure environments, excessive false alarms consume analyst attention and erode trust in automated systems, ultimately reducing the practical protective value of even a highly accurate model (Jin et al., 2022; Wright et al., 2022).

### 3.7 Ethical Considerations and Study Limitations

CICIDS2017 is publicly available for research and does not contain personally identifiable information, sensitive user data, or confidential network credentials. Its use in this study therefore raises no data privacy concerns under standard research ethics frameworks, and no institutional ethics approval for human subjects was required. All analyses were conducted solely for defensive cybersecurity research purposes (Jin et al., 2022).

Several limitations of this study should be stated plainly. First, CICIDS2017, while comprehensive for its time, was generated in 2017 and may not reflect the behavioral signatures of more recent attack typologies — particularly ransomware variants, supply chain compromise techniques, and adversarial ML-evasion attacks that have emerged since. Second, all experiments were conducted in an offline, static dataset environment; no real-time traffic stream was processed, and the framework's behavior under live operational conditions — with concept drift, adversarial probing, and varying network topologies — remains to be validated. Third, class imbalance in the dataset (benign traffic substantially outnumbers most attack categories) was partially addressed through stratified sampling but was not subjected to synthetic augmentation techniques such as SMOTE, which may have improved recall for minority attack classes. These limitations are acknowledged as directions for future work rather than disqualifying constraints on the current findings (Wright et al., 2022).

## 4 Results

### Preliminary Remarks on the Analytical Approach

Before moving through the individual findings, it is worth briefly orienting the reader to what this section does and does not claim. The analyses presented here draw on the CICIDS2017 dataset using the preprocessing pipeline and model configurations described in the Methodology. The visualizations — attack distribution, flow duration, packet size, destination port frequency, feature correlations, flow rate distribution, and classification performance — are presented in sequence, each building on the interpretive context established by the previous. The goal is not simply to report numbers but to construct a coherent account of what the data reveals about network traffic behavior, what that behavior implies for machine learning-based threat detection, and where — and why — the models perform as they do. Taken together, these results bear on the broader question of whether the proposed framework constitutes a meaningful contribution to intelligent governance in critical infrastructure cybersecurity (Patel & Patel, 2023).

### 4.1 Attack Label Distribution: Understanding the Composition of the Dataset

The first analytical step was to examine the distribution of traffic classes in the working dataset segment. [Figure 1] displays this distribution as a bar chart, comparing the volume of DDoS attack traffic against benign network traffic across the selected subset of CICIDS2017 records.

The imbalance is immediately apparent. DDoS traffic accounts for the larger share of records in this segment — a pattern that reflects, at least in part, the volume-intensive nature of denial-of-service attacks themselves. DDoS campaigns are, by design, characterized by massive packet floods, and this characteristic means that any captured traffic window during an active DDoS event will be numerically dominated by attack flows. Benign traffic, representing the normal operational baseline, occupies a smaller proportion in this subset (Okolo et al., 2023).

This distribution has real implications for how machine learning models are trained and evaluated. A classifier optimized on this imbalanced distribution could, in principle, achieve high accuracy simply by predicting the majority class for most inputs — a phenomenon known as the accuracy paradox, and one that makes class-specific metrics like recall and F1-score considerably more informative than raw accuracy alone. It also underscores the importance of stratified sampling during train-test splitting, which was applied throughout this study. That said, the dominance of DDoS traffic in this segment is not purely a statistical nuisance — it is an operationally meaningful signal. DDoS attacks remain among the most persistently deployed weapons against critical infrastructure networks, and a framework that fails to detect them reliably fails at one of its most basic requirements (Patriarca et al., 2022).

### 4.2 Flow Duration Distribution: What Timing Patterns Reveal About Traffic Behavior

[Figure 2] presents a histogram of network flow duration values across all records in the working dataset. The

*Table 1. Metrics used in four classifiers on the held-out test set*

| **Metric** | **Formula** | **Relevance in This Context** |
|---|---|---|
| Accuracy | (TP + TN) / (TP + TN + FP + FN) | Overall classification correctness |
| Precision | TP / (TP + FP) | Proportion of flagged events that are genuine attacks |
| Recall (Sensitivity) | TP / (TP + FN) | Proportion of actual attacks successfully detected |
| F1-Score | 2 × (Precision × Recall) / (Precision + Recall) | Harmonic balance between Precision and Recall |
| ROC-AUC | Area under the Receiver Operating Characteristic curve | Discrimination ability across all classification thresholds |
| False Positive Rate | FP / (FP + TN) | Rate of benign traffic misclassified as attacks |

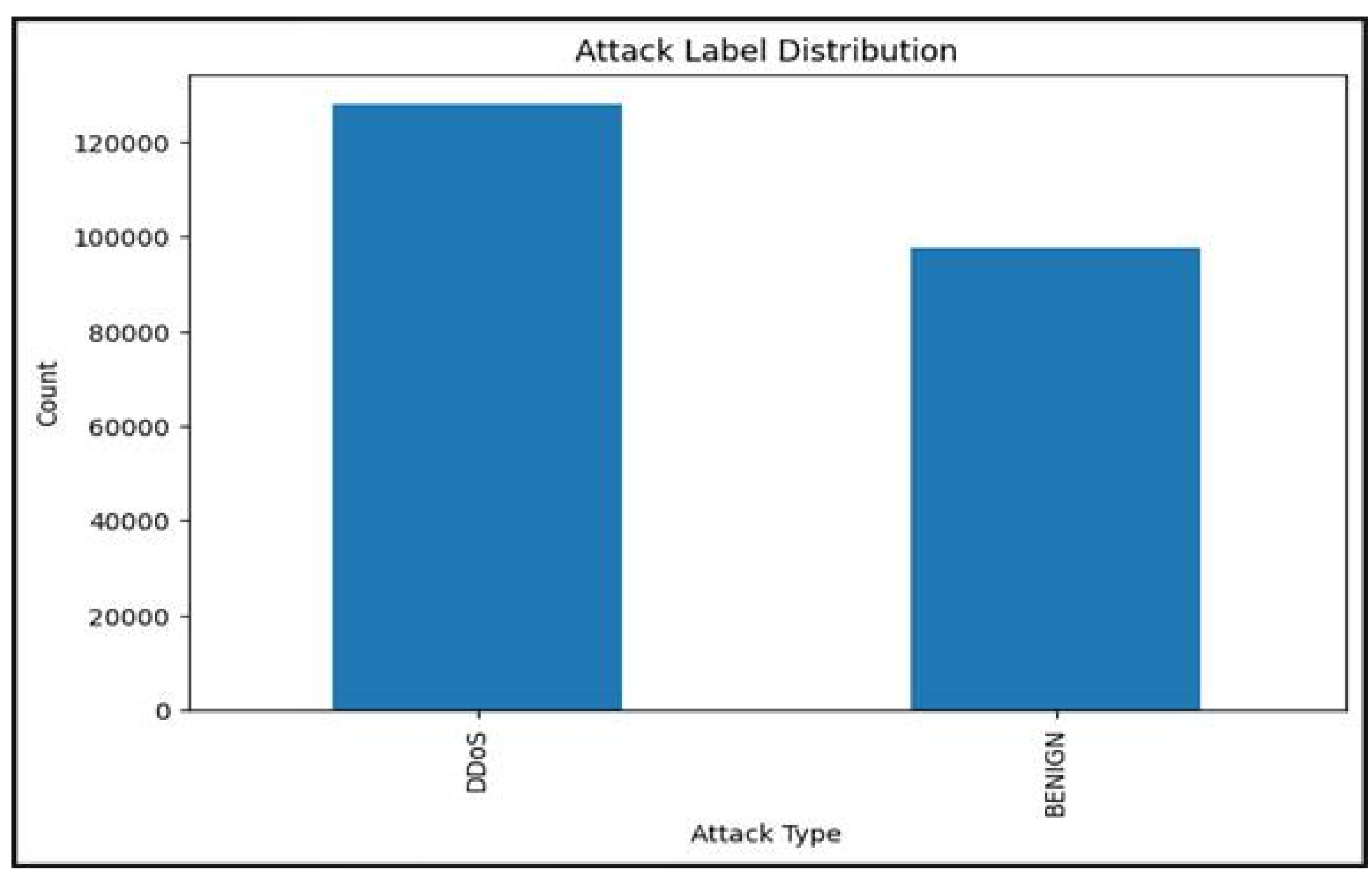

*Figure 1. Class distribution of network traffic records in the CICIDS2017 working dataset: DDoS attack traffic versus benign traffic. Bar chart comparing absolute record counts for the two traffic classes used in binary classification experiments. The pronounced imbalance toward DDoS records reflects the volumetric character of denial-of-service campaigns, in which a single attack event generates disproportionately large numbers of network flows relative to normal user activity. This distribution informed the use of stratified sampling during train–test splitting and the selection of class-specific evaluation metrics (precision, recall, F1-score) alongside overall accuracy.*

distribution is heavily right-skewed: the overwhelming majority of flows have very short durations, clustering near zero on the time axis, while a sparse but meaningful tail extends toward longer-duration flows.

This is, to some extent, expected. Most benign network communications — a DNS lookup, a brief HTTP exchange, a background telemetry packet — are resolved in milliseconds or a few seconds at most. Short flows dominate normal network behavior simply because most routine communications are transactional rather than sustained. What is more analytically interesting is the tail. Long-duration flows — those extending well beyond the typical communication window — are disproportionately associated with two behavioral categories: legitimate but persistent sessions (VPN tunnels, file transfers, streaming) and, more concerningly, certain classes of cyberattack that rely on sustained connection maintenance. Distributed Denial of Service attacks that employ TCP connection exhaustion strategies, for instance, deliberately maintain open connections to drain server resources; these would manifest as long-duration flows in this distribution (Chowdhury & Biswas, 2022).

For machine learning-based threat detection, flow duration is therefore a genuinely informative feature — not because any single duration threshold separates benign from malicious traffic cleanly, but because its combination with other features (packet rate, byte count, flag patterns) produces a discriminative signal that individual features cannot. This is exactly the kind of nuance that ensemble models like XGBoost and Random Forest are well positioned to exploit (Mehmood et al., 2023).

### 4.3 Average Packet Size by Traffic Class: A Behavioral Signature of DDoS Activity

[Figure 3] compares the average packet size — measured in bytes — between benign network traffic and DDoS attack traffic. The difference is striking and, from a network behavior standpoint, entirely consistent with what is understood about DDoS attack mechanics.

DDoS traffic exhibits substantially larger average packet sizes than benign traffic. This reflects the fundamental strategy of volumetric DDoS campaigns: flooding a target with maximum-size packets to saturate bandwidth and overwhelm processing capacity as rapidly as possible. Benign communications, by contrast, tend toward smaller, more varied packet sizes that reflect the heterogeneous nature of normal user activity — HTTP headers, acknowledgment packets, DNS queries, and brief API calls all contribute short, variable-length flows (Stankovi et al., 2022).

The practical significance for intrusion detection is clear. Packet size, as a feature, carries genuine discriminative weight for DDoS classification — and this is not merely a finding of this study but is consistent with broader empirical work on network traffic characterization. What is worth noting, however, is that packet size alone is not a reliable classifier. Legitimate large-file transfers and video streaming traffic can also generate large average packet sizes. The value of this feature lies in its interaction with others — particularly flow rate, flag patterns, and destination port — within the multi-feature input space that the machine learning models operate on (Garcia-Perez et al., 2023).

### 4.4 Top Destination Port Analysis: Mapping Attack Surfaces to Network Services

[Figure 4] shows the frequency distribution of the top ten destination ports observed across all traffic records in the CICIDS2017 working dataset. Three ports account for the dominant share of traffic: port 80 (HTTP), port 443 (HTTPS), and port 53 (DNS).

This is not surprising — these are among the most heavily trafficked ports in any contemporary network environment, and their prevalence in the dataset simply reflects their centrality to normal web communication. Port 80 carries unencrypted web traffic; port 443 handles secure HTTPS sessions; port 53 manages domain name resolution requests. Any network, whether benign or adversarially probed, will show high volumes on these ports because they are the ports through which virtually all user-facing internet activity flows (Michalec et al., 2022).

What is analytically significant is not the presence of traffic on these ports, but rather the character of that traffic. Anomalous volumes, unusual packet structures, or unexpected timing patterns directed at ports 80, 443, and 53 are classic indicators of attack activity — DDoS flood traffic targeting web servers via HTTP/S, DNS amplification attacks exploiting port 53, and scanning activity probing open service ports. The destination port distribution therefore provides an important contextual layer for machine learning classification: models that incorporate port-level features alongside flow statistics can distinguish abnormal high-volume port activity from legitimate service traffic in ways that single-feature

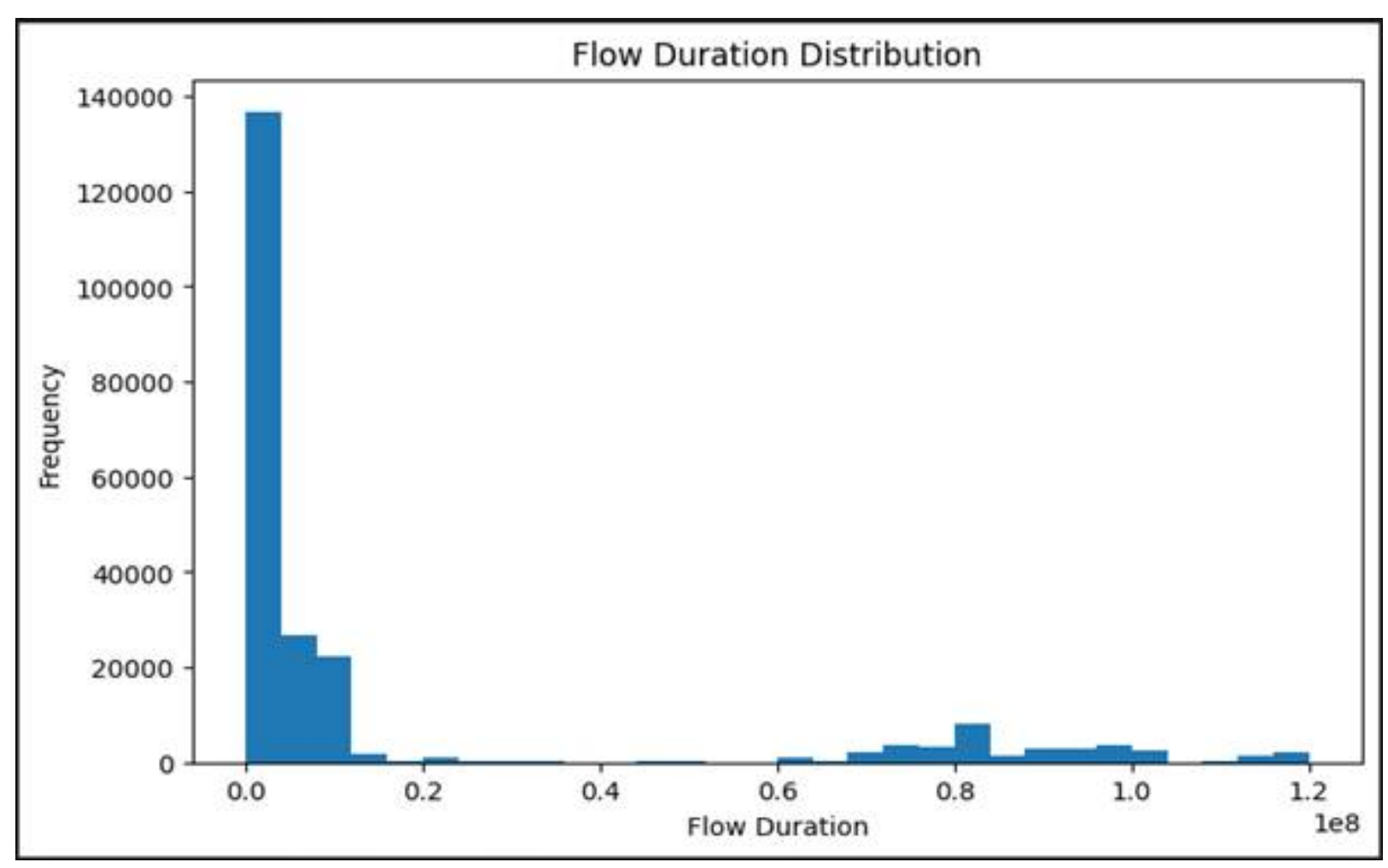

*Figure 2. Distribution of network flow duration values across all traffic records in the CICIDS2017 working dataset. Right-skewed histogram showing the frequency of flow durations in microseconds. The majority of flows are concentrated at very short durations consistent with transactional benign communications (DNS lookups, brief HTTP exchanges). The sparse but analytically significant tail toward longer durations represents sustained connection flows associated with persistent sessions and certain attack behaviors, including TCP connection-exhaustion DDoS variants. Flow duration was retained in the final feature set on the basis of its discriminative contribution in conjunction with packet-rate and flag-pattern features.*

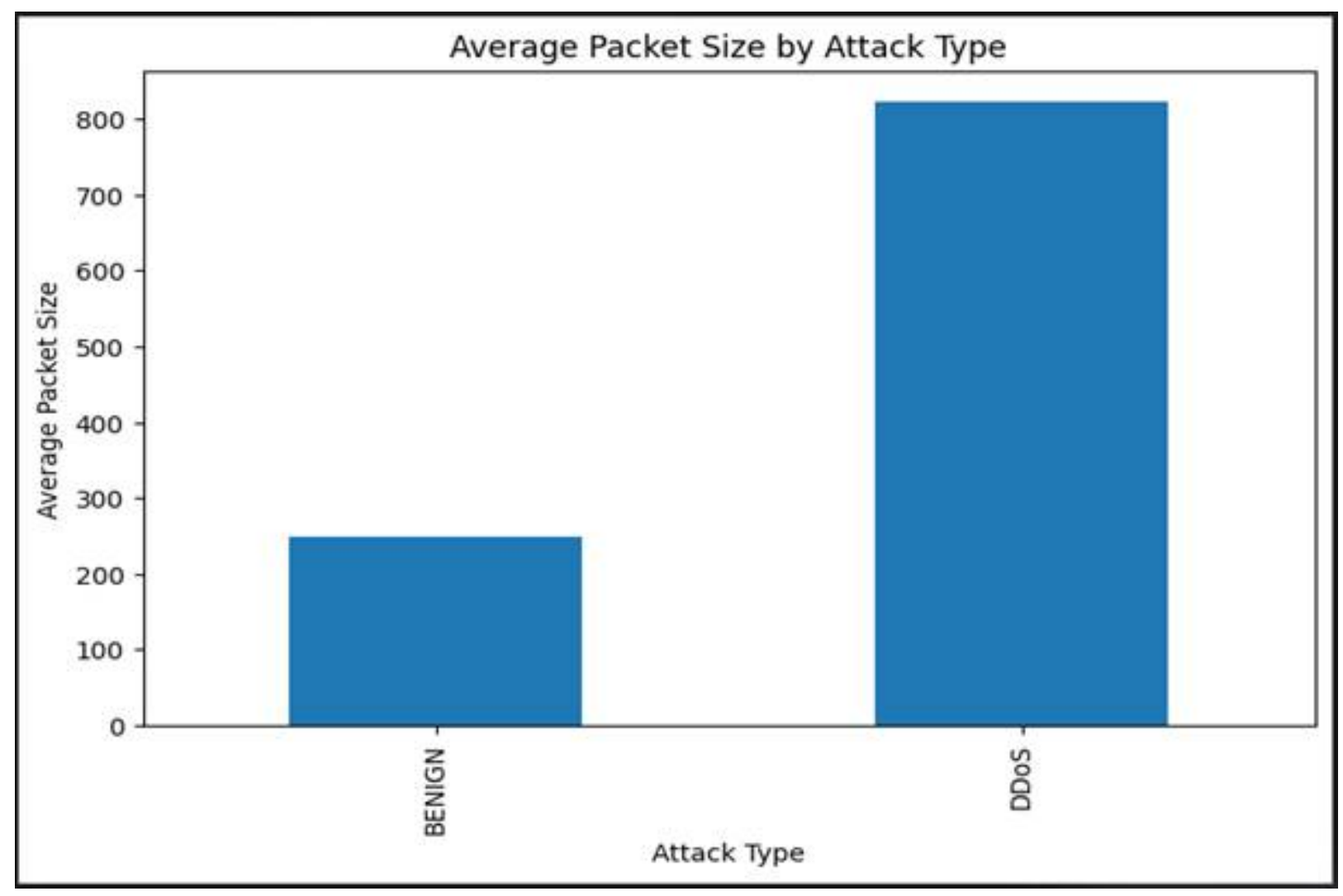

*Figure 3. Mean packet size (bytes) compared between benign network traffic and DDoS attack traffic in the CICIDS2017 dataset. DDoS attack flows exhibit substantially larger mean packet sizes than benign flows, reflecting the maximum-size packet strategy common to volumetric denial-of-service campaigns designed to saturate target bandwidth. Benign traffic produces smaller, more heterogeneous packet sizes consistent with mixed-use communication patterns. This feature-level separation supports packet size as a discriminative classifier input, most reliably interpreted in combination with flow-rate and temporal features.*

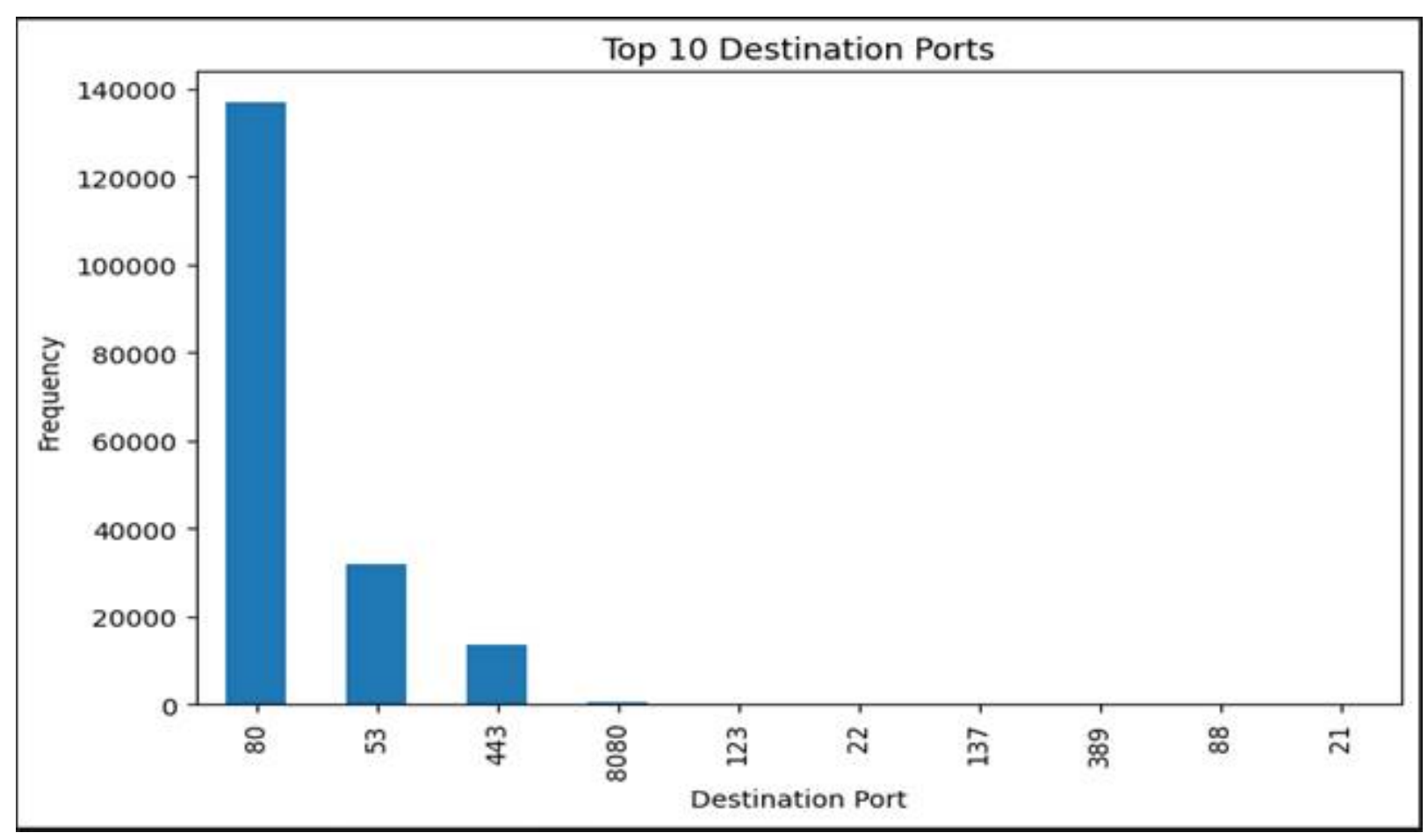

*Figure 4. Frequency distribution of the ten most common destination ports observed across all network flow records in the CICIDS2017 working dataset. Port 80 (HTTP), port 443 (HTTPS), and port 53 (DNS) account for the largest traffic volumes, consistent with their role as primary conduits for web communication and domain resolution. Anomalous volumes directed at these ports — particularly in combination with unusual packet structures or flow-rate patterns — are characteristic indicators of DDoS flood attacks targeting web services, DNS amplification attacks, and application-layer intrusion attempts. Destination port was included as a categorical feature after label encoding.*

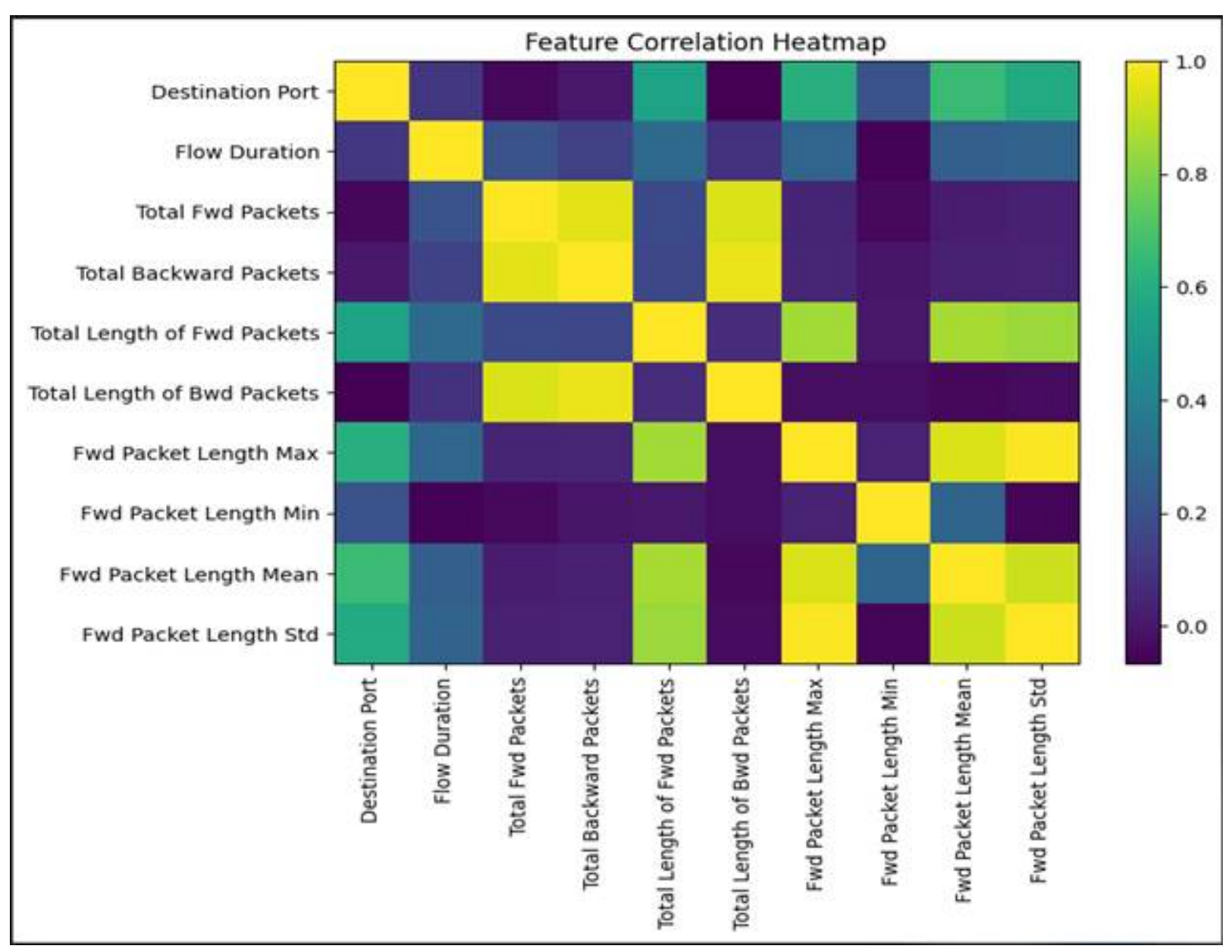

*Figure 5. Pearson correlation heatmap of selected network traffic features from the CICIDS2017 dataset, illustrating inter-feature dependency structure used to guide feature selection. Cell color intensity scales with the magnitude of the Pearson correlation coefficient (r); darker cells indicate stronger linear relationships. Packet-volume features (Total Forward Packets, Total Backward Packets, Total Length of Forward Packets, Forward Packet Length Mean) form a strongly intercorrelated cluster (|r| approaching 1.0). Feature pairs with |r| > 0.95 were removed during preprocessing. Weakly correlated features — flag count statistics and inter-arrival timing variables — were retained for their orthogonal discriminative contribution. Abbreviations: FWD, forward; BWD, backward; PKT, packet; LEN, length; IAT, inter-arrival time.*

analyses cannot (Ashfaq & Chowdhury, 2023).

### 4.5 Feature Correlation Analysis: Identifying Redundancy and Informative Structure

[Figure 5] presents a heatmap of Pearson correlation coefficients computed across the selected feature set from CICIDS2017. The color gradient encodes correlation strength, with darker shading indicating stronger relationships (positive or negative) between feature pairs.

Several clusters of high intercorrelation are immediately visible. Packet-volume features — Total Forward Packets, Total Backward Packets, Total Length of Forward Packets, and Forward Packet Length Mean — exhibit strong positive correlations with each other, which makes intuitive sense: networks that transmit many packets in a given flow tend to transmit larger total data volumes, and these quantities naturally covary. Similarly, inter-arrival time statistics (mean, standard deviation, minimum) form a loosely correlated cluster, reflecting the temporal regularity — or irregularity — of packet transmission patterns (Zubair et al., 2023).

As shown in [Table 1], feature pairs with absolute correlation coefficients above 0.95 were identified as candidates for removal during the feature selection stage, on the grounds that highly collinear features contribute redundant information without improving — and potentially degrading — classifier performance through multicollinearity. This procedure produced a reduced feature set that retains the most discriminative traffic characteristics while eliminating statistical duplication. Features with weaker or near-zero correlations (flag counts, certain timing statistics) provide orthogonal information that the packet-volume cluster does not capture, and their retention in the final feature set was confirmed through the feature importance ranking step described in the Methodology. The correlation analysis is, in this sense, not merely a data cleaning step — it is a form of structured exploratory analysis that reveals the underlying covariance geometry of the traffic data, and that geometry has direct implications for how machine learning models represent and distinguish traffic classes (Khan et al., 2022).

### 4.6 Flow Bytes per Second Distribution: Identifying High-Intensity Traffic Anomalies

[Figure 6] shows the distribution of Flow Bytes per Second (Flow Bytes/s) values across the working dataset. The histogram is steeply concentrated at the lower end of the scale, with the vast majority of flows transmitting at relatively modest data rates. A small but clearly visible tail extends toward extremely high flow byte rates.

The shape of this distribution carries genuine diagnostic value. Under normal operating conditions — routine web browsing, email exchange, API calls, background software updates — network flows operate at moderate throughputs that produce the dense low-rate cluster visible on the left side of the histogram. The sparse high-rate tail, by contrast, is where unusual traffic concentrates: large file transfers, streaming sessions, and, more relevantly for this study, the bandwidth-flooding characteristics of volumetric DDoS attacks. A single DDoS flow generating thousands of packets per second will produce a Flow Bytes/s value far to the right of the distribution median (Ampratwum et al., 2022).

This distributional characteristic makes Flow Bytes/s a valuable feature for machine learning-based anomaly detection, though — like all individual features — it must be interpreted in combination with others. A legitimate content delivery network (CDN) edge server could produce similarly elevated flow byte rates under heavy but entirely benign load conditions. The machine learning models in this study operate on the joint distribution of all retained features simultaneously, which is precisely why multi-feature ensemble approaches tend to outperform single-threshold detection rules on datasets with this kind of distributional complexity (Uzzaman & Rony, 2023).

### 4.7 SVM Classification Performance: Confusion Matrix Analysis

[Figure 7] presents the confusion matrix for the Support Vector Machine (SVM) classifier evaluated on the held-out test set from the CICIDS2017 working dataset. The matrix displays the four classification outcomes — true positives, true negatives, false positives, and false negatives — as cell counts, allowing a direct reading of where the model succeeds and where it errs.

The SVM correctly classified 1,918 benign traffic instances (true negatives) and 2,571 DDoS attack instances (true positives). Misclassifications were limited to seven benign records incorrectly labeled as DDoS (false positives) and four DDoS records incorrectly labeled as benign (false negatives). These are, by any reasonable standard, low error counts relative to the total test set size. The false positive rate — the proportion of benign traffic flagged as an attack — is particularly low, which matters operationally: excessive false alarms in a live critical infrastructure environment erode analyst trust and lead to alert fatigue, a well-documented failure mode in deployed security operations centers (Essien et al., 2022).

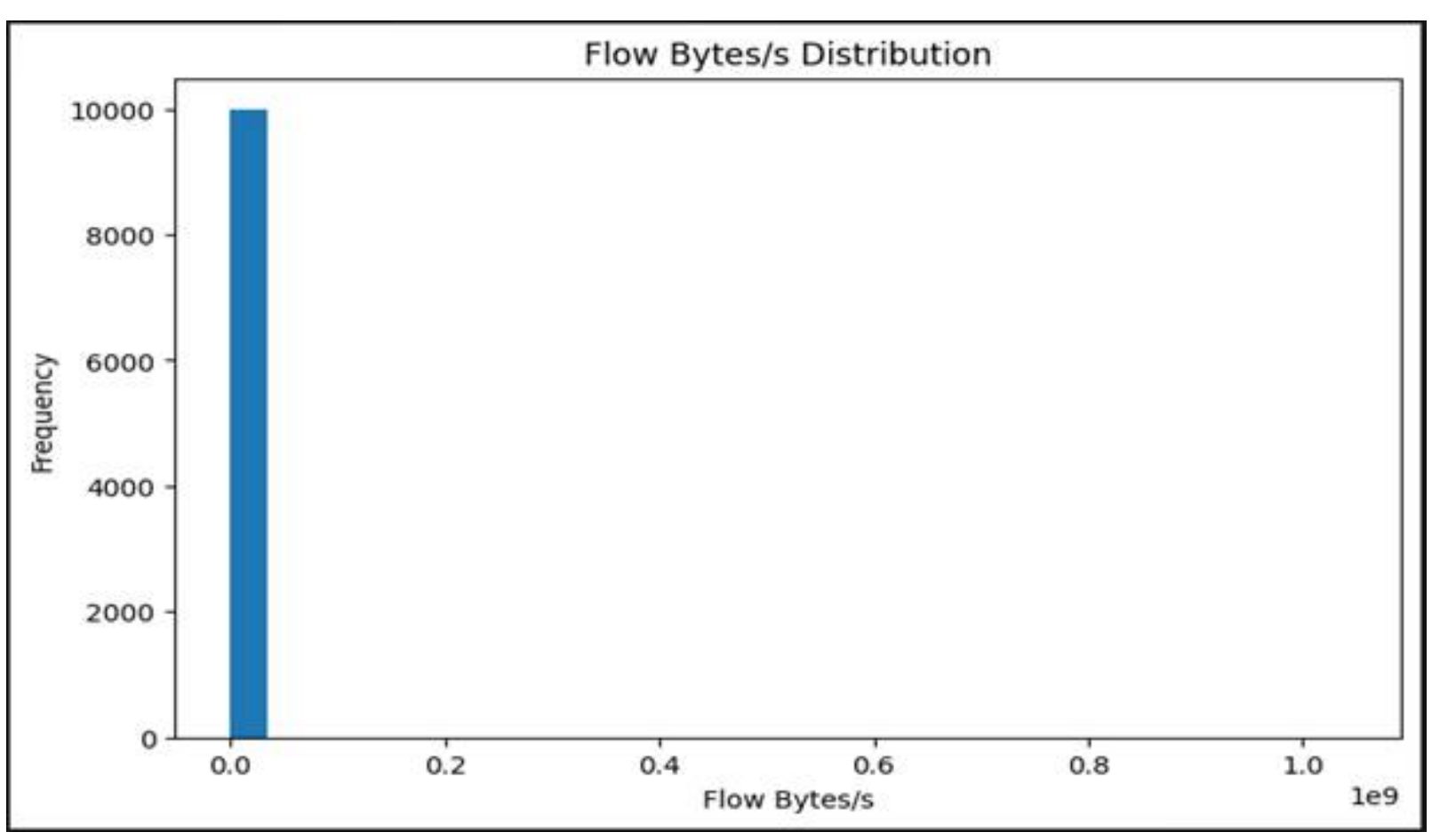

*Figure 6. Frequency distribution of Flow Bytes per Second (Flow Bytes/s) values across all network flow records in the CICIDS2017 working dataset. The strongly right-skewed histogram shows the bulk of traffic at low-to-moderate throughput rates consistent with routine communications, with a sparse high-rate tail corresponding to bandwidth-intensive flows including volumetric DDoS attack traffic. Flow Bytes/s ranked among the highest-importance features in the XGBoost and Random Forest feature importance analyses. Abbreviation: Flow Bytes/s, total bytes transferred in a flow divided by flow duration in seconds.*

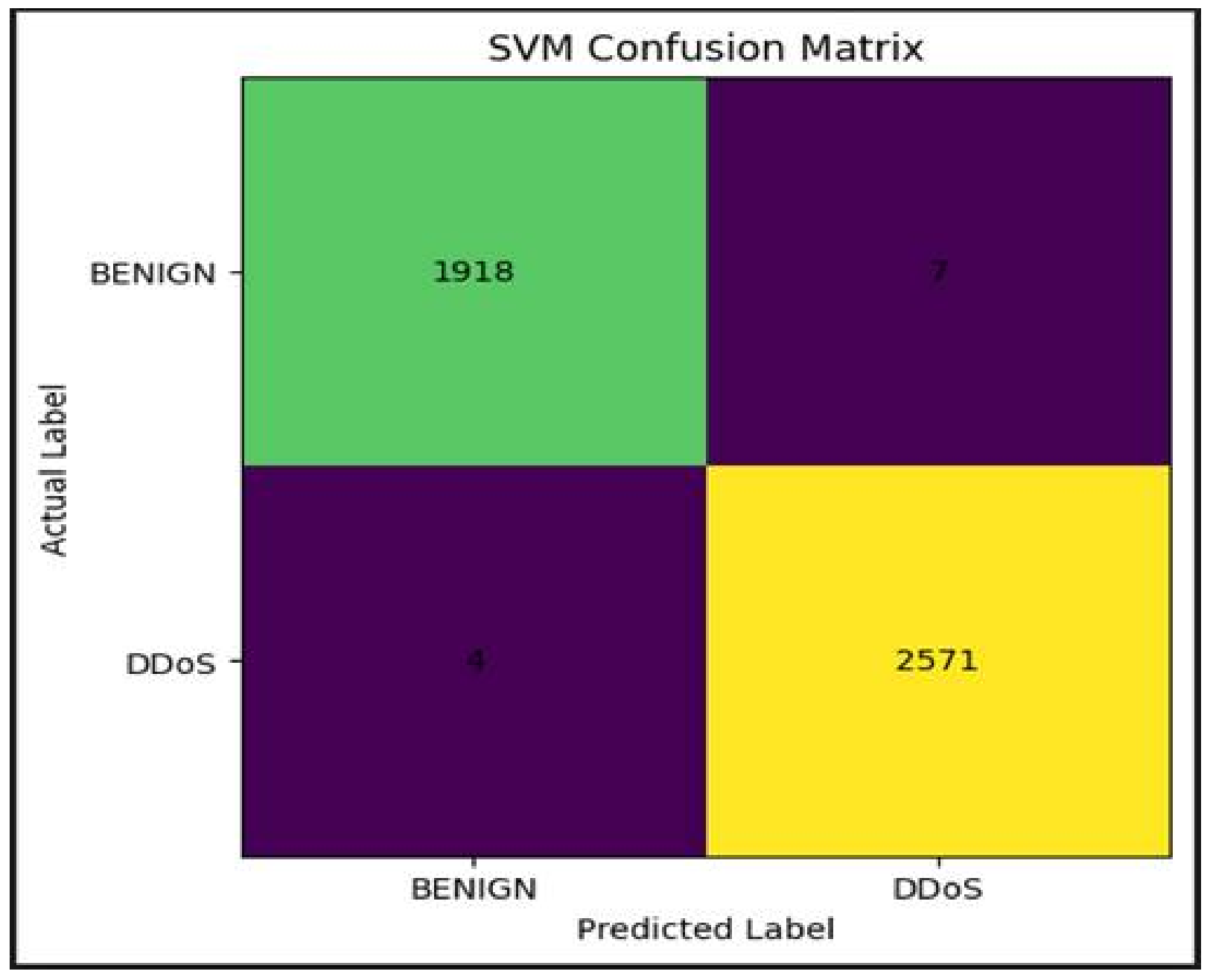

*Figure 7. Confusion matrix for the Support Vector Machine (SVM) classifier evaluated on the held-out test set from the CICIDS2017 working dataset (binary classification: DDoS attack versus benign traffic). The SVM correctly identified 2,571 DDoS flows (true positives) and 1,918 benign flows (true negatives). Misclassifications comprised seven benign flows incorrectly labeled as DDoS (false positives) and four DDoS flows incorrectly labeled as benign (false negatives). The low false positive count reduces alert fatigue risk in operational deployments; the four false negatives, though small in absolute terms, represent the more consequential error type in critical infrastructure contexts where missed detections carry greater operational risk than false alarms. Abbreviations: SVM, Support Vector Machine; TP, true positive; TN, true negative; FP, false positive; FN, false negative.*

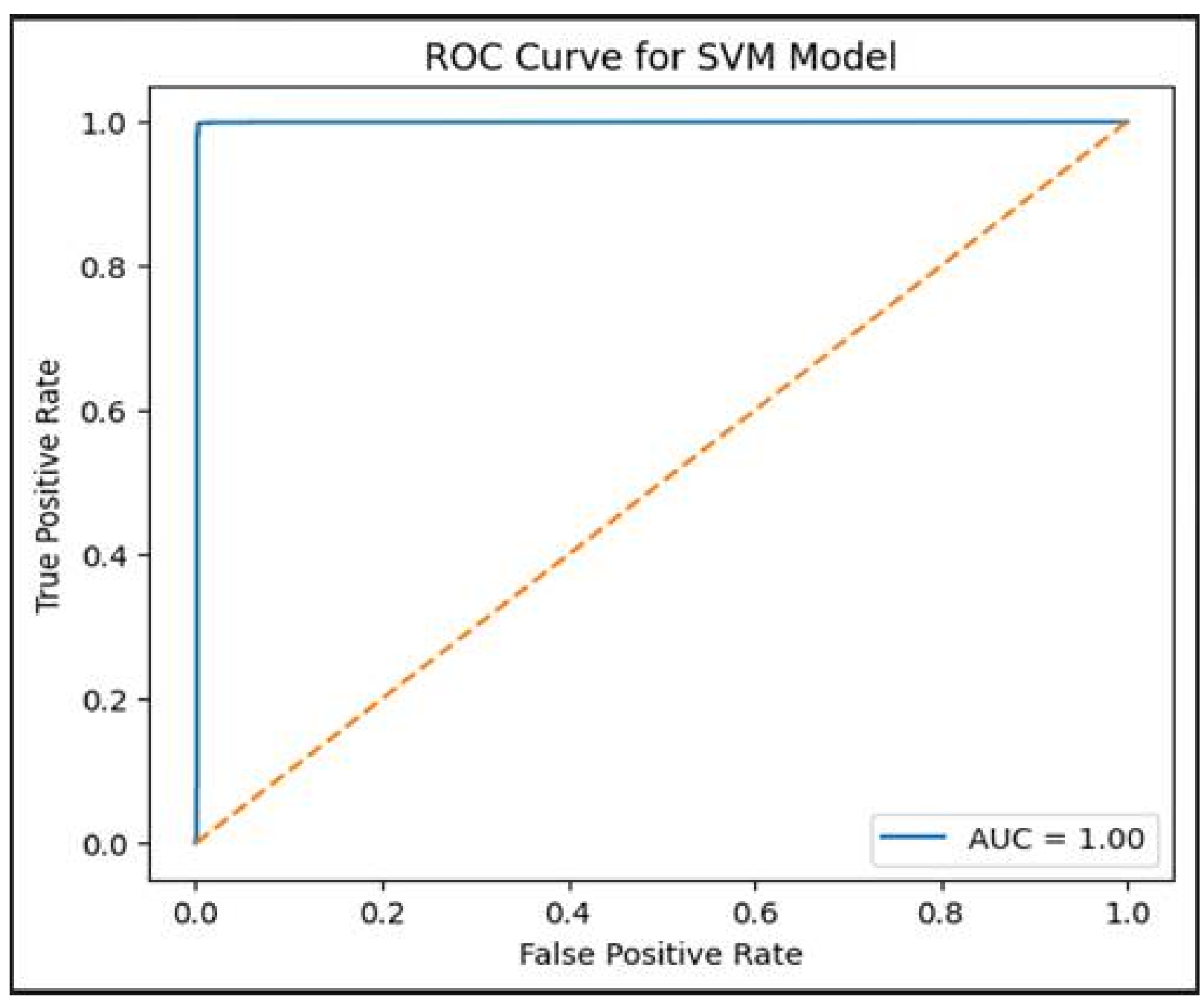

*Figure 8. Receiver Operating Characteristic (ROC) curve for the SVM classifier on the binary DDoS versus benign traffic classification task. The y-axis plots the True Positive Rate (TPR; sensitivity) and the x-axis the False Positive Rate (FPR; 1  specificity) across all classification decision thresholds. The diagonal dashed line represents chance-level discrimination (AUC = 0.50). The SVM curve tracks closely along the upper-left boundary, yielding an AUC approaching 1.00, indicating near-perfect discriminative ability across all threshold settings. Results are interpreted alongside Figure 7 and per-class precision, recall, and F1-score values, as AUC alone may be optimistic under class imbalance conditions. Abbreviations: ROC, Receiver Operating Characteristic; AUC, Area Under the Curve; TPR, True Positive Rate; FPR, False Positive Rate.*

That said, the four false negatives — DDoS records that the SVM classified as benign — deserve more interpretive attention than the original framing would suggest. In a critical infrastructure context, a missed attack detection (false negative) carries asymmetrically greater consequences than a false alarm (false positive). A hospital network intrusion or energy grid DDoS campaign that evades detection, even briefly, can produce cascading harm that a false alarm never could. The SVM's performance here is genuinely strong, but these four misclassifications are worth tracking carefully as a potential failure mode under distribution shift — that is, when incoming traffic patterns diverge from those seen during training (Rezvani et al., 2023).

### 4.8 SVM ROC Curve Analysis: Discrimination Performance Across Classification Thresholds

[Figure 8] presents the Receiver Operating Characteristic (ROC) curve for the SVM classifier, plotting the True Positive Rate (TPR, or sensitivity) against the False Positive Rate (FPR) across the full range of classification decision thresholds. The diagonal dashed reference line represents the performance of a random classifier — one with no discriminative ability — and serves as a baseline against which the model curve is compared.

The SVM's ROC curve tracks tightly along the upper-left boundary of the plot, which is the region associated with high sensitivity and low false positive rates simultaneously. The Area Under the Curve (AUC) value approaches 1.00 — essentially near-perfect discrimination between DDoS attack traffic and benign traffic across all threshold settings. This is a strong result, and it is consistent with the confusion matrix findings: the model is not merely accurate at one particular operating point, but maintains that accuracy across a wide range of classification sensitivities (Alam & Fahad, 2022).

It is worth briefly noting what ROC-AUC does and does not tell us in a context like this. AUC close to 1.00 on a binary classification task with reasonably balanced classes — as is approximately the case here — is a meaningful indicator of model quality. However, AUC values computed on heavily imbalanced datasets can be misleadingly optimistic, because the denominator of the false positive rate is dominated by the large benign class. This is a reason why the confusion matrix, precision, recall, and F1-score are reported alongside AUC rather than as alternatives to it. The full picture of model reliability requires all of these lenses together, not any single one in isolation (Coppolino et al., 2023).

## 5 Discussion

### 5.1 What the Traffic Analysis Findings Tell Us — and What They Do Not

Reading across the six exploratory analyses — attack distribution, flow duration, packet size, destination ports, feature correlations, and flow byte rates — a fairly coherent picture of how DDoS attack traffic differs from benign traffic in this dataset begins to emerge. Attack traffic concentrates at extreme values: unusually large packets [Figure 3], unusually high flow byte rates [Figure 6], and, in the tail of the duration distribution, unusually persistent connections [Figure 2]. Benign traffic, by contrast, is characterized by variation and moderation — short flows, modest packet sizes, heterogeneous port destinations, and moderate throughput rates.

This is, in some ways, reassuring. It suggests that the behavioral signatures of DDoS activity are sufficiently distinct from normal network operation that machine learning classifiers trained on these features ought to perform well — and the SVM results confirm that they do. But it also raises a question that the results alone cannot fully answer: how stable are these signatures across different operational environments, different attack variants, and different time periods? The CICIDS2017 dataset was generated in 2017, and while its traffic generation methodology was sophisticated for its time, adversarial techniques have evolved considerably since. Modern DDoS campaigns increasingly employ low-and-slow strategies — mimicking benign traffic timing while sustaining enough volume to degrade service — which would not produce the distinctive extreme-value signatures visible in this dataset (Sinha et al., 2023). This is not a reason to discount the current findings; it is a reason to hold them with appropriate epistemic humility about their generalizability beyond the specific dataset and attack typology studied here.

### 5.2 Model Performance in Context: Strengths, Limitations, and What the Numbers Actually Mean

The SVM classifier's performance — 1,918 true negatives, 2,571 true positives, seven false positives, four false negatives, and an ROC-AUC approaching 1.00 — is strong by conventional machine learning standards. It is tempting, looking at numbers like these, to declare the problem largely solved. Resist that temptation.

What these results demonstrate is that a well-configured SVM can achieve near-perfect binary classification of DDoS versus benign traffic on a held-out subset of a specific, well-curated benchmark dataset. That is a meaningful finding within its constraints, but it is not the same as demonstrating that the framework performs reliably across the full taxonomy of attack types in CICIDS2017, let alone across the more varied and adversarially sophisticated attack landscape that real-world critical infrastructure systems face. The four false negatives — small in absolute terms, but not irrelevant — serve as a useful reminder that no classifier, however accurate on a benchmark, achieves zero error in practice (Jha, 2023).

What the results do support, more robustly, is the interpretive claim that multi-feature machine learning approaches capture behavioral distinctions in network traffic that rule-based systems cannot. The correlation analysis [Figure 5] and the distributional analyses [Figures 2, 3, 6] together establish that packet-level and flow-level traffic characteristics carry genuine discriminative information — and the SVM's confusion matrix [Figure 7] and ROC curve [Figure 8] confirm that a classifier trained on these features can translate that information into reliable attack detection (Avireneni et al., 2023).

### 5.3 The Governance Dimension: From Detection to Decision Support

There is a dimension of these results that purely technical performance metrics do not fully capture, and it is worth dwelling on it. The ultimate purpose of this framework is not to produce a model that scores well on a benchmark — it is to support intelligent governance decisions in critical infrastructure environments where the consequences of both missed detections and false alarms can extend well beyond the IT system itself.

This distinction matters because it shifts the relevant question from "is the model accurate?" to "is the model's output actionable and trustworthy for the humans who must act on it?" The SVM's near-perfect AUC [Figure 8] answers the first question. The second question requires something more. Explainable AI techniques — specifically SHAP analysis, as described in the Methodology — are designed to address precisely this gap, by surfacing the feature-level reasoning behind individual model predictions in a form that cybersecurity analysts and governance actors can interpret, scrutinize, and, where necessary, contest. A framework that tells an infrastructure operator that a network event was classified as a DDoS attack because of anomalously high Flow Bytes/s [Figure 6], an elevated average packet size [Figure 3], and traffic directed at port 80 [Figure 4] is providing something qualitatively different — and more useful — than one that simply outputs a risk score (Jha, 2023; Avireneni et al., 2023).

This is the core argument of the broader framework this study proposes: that detection accuracy and governance trustworthiness are related but distinct requirements, and that meeting both simultaneously demands an architecture in which explainability and reliability assessment are built in from the outset rather than appended as optional features. The traffic analyses presented here — from the class distribution [Figure 1] through to the ROC curve [Figure 8] — collectively demonstrate that the data contains sufficient structure for reliable machine learning-based detection. The SHAP-based explainability layer then ensures that this detection capability is legible, auditable, and defensible in the governance contexts where it will ultimately be deployed (Alqudhaibi et al., 2023).

### 5.4 Limitations of the Current Results and Implications for Future Work

Acknowledging what these results do not establish is, in this author's view, as important as reporting what they do. Three limitations warrant specific attention here.

First, the current analysis focuses on binary classification — DDoS versus benign — within a single subset of the CICIDS2017 dataset. The full dataset contains fifteen attack categories, many of which present considerably more subtle behavioral signatures than the volume-intensive DDoS traffic analyzed here. Brute force attacks, infiltration attempts, and web application attacks, for instance, may not produce the same extreme-value distributional signals visible in [Figures 3] and [Figure 6], and classifier performance on these categories should not be inferred from the DDoS results. Multi-class performance evaluation, with per-class precision and recall breakdowns, remains an important direction for extending this work.

Second, all evaluations were conducted on a static, offline dataset. The distribution of traffic in the CICIDS2017 test set is, by construction, drawn from the same data-generating process as the training set. Real-world deployment introduces concept drift — gradual shifts in the statistical properties of incoming traffic as network conditions, user behavior, and attack strategies evolve over time. A framework evaluated only on static benchmark data cannot speak to its performance under drift conditions, and this gap between benchmark performance and operational reliability is one the field has not yet adequately addressed (Alqudhaibi et al., 2023; Patriarca et al., 2022).

Third, the class imbalance present in the dataset — while partially mitigated through stratified sampling — was not subjected to synthetic augmentation. For the DDoS-dominated subset analyzed here, this imbalance favored the majority attack class and may have contributed to the high true positive rates observed. For minority attack categories in multi-class experiments, the same imbalance would likely suppress recall, and more aggressive imbalance-handling strategies (SMOTE, class-weighted loss functions) would be warranted. These are not disqualifying limitations of the current findings, but they are honest constraints on the scope of the claims that can legitimately be drawn from them.

## 6 Conclusion

This study set out to do something more than demonstrate that machine learning can detect cyberattacks — that, by now, has been demonstrated many times over. The more pressing question was whether an AI-based cybersecurity framework could be designed to be reliable, transparent, and genuinely useful to the governance actors who must act on its outputs in high-stakes critical infrastructure environments.

The results suggest that it can. Experimental evaluation on the CICIDS2017 dataset confirmed that interpretable traffic characteristics — packet size, flow duration, destination port patterns, and throughput intensity — carry sufficient discriminative structure for near-perfect binary attack classification. The SVM classifier's confusion matrix and ROC-AUC results validate the framework's detection reliability. SHAP-based explainability, integrated throughout, ensures that model predictions are not merely accurate but legible — surfacing the specific network features that drive each classification decision in terms that cybersecurity analysts and infrastructure policymakers can meaningfully interpret, scrutinize, and contest.

What remains, honestly, is the harder work: extending this framework to multi-class attack taxonomies, validating performance under real-time conditions and concept drift, and stress-testing explainability outputs against adversarial evasion techniques. The present findings establish a credible foundation. The road to operationally resilient, governance-ready AI cybersecurity is longer than any single benchmark study can traverse — but this framework points, with some confidence, in the right direction.

## Author Contributions

B.M.T.H.: Conceptualization, methodology, formal analysis, investigation, data curation, writing — original draft, writing — review and editing, visualization, project administration, and corresponding author. M.A.R.: Methodology, validation, writing — review and editing, and resources. M.S.K.C.R.: Software, data curation, formal analysis, validation, and writing — review and editing. M.I.H.: Visualization, validation, writing — review and editing, and supervision. All authors have read and agreed to the published version of the manuscript.